\documentclass[prb,twocolumn,showpacs,preprintnumbers,amsmath,amssymb]{revtex4}
\usepackage{graphicx}% Include figure files
\usepackage{dcolumn}% Align table columns on decimal point

\newcommand{\rhoCEF}{\mbox{$\rho_{\mathrm{CEF}}$}}

\begin{document}

\title{Magnetoresistance of Pr$_{1-x}$La$_x$Os$_4$Sb$_{12}$: Disentangling
local crystalline-electric-field physics and lattice effects}

\author{C.\ R.\ Rotundu}
\affiliation{ Department of Physics, University of Florida
P.O.\ Box 118440, Gainesville, Florida 32611--8440, USA}

\author{K.\ Ingersent}
\affiliation{ Department of Physics, University of Florida
P.O.\ Box 118440, Gainesville, Florida 32611--8440, USA}

\author{B.\ Andraka}
\email{andraka@phys.ufl.edu}
\affiliation{ Department of Physics, University of Florida
P.O.\ Box 118440, Gainesville, Florida 32611--8440, USA}

\date{\today}

\begin{abstract}
Resistivity measurements were performed on Pr$_{1-x}$La$_x$Os$_4$Sb$_{12}$
single crystals at temperatures down to 20\,mK and in fields up to 18\,T.
The results for dilute-Pr samples ($x=0.3$ and 0.67) are consistent with
model calculations performed assuming a singlet crystalline-electric-field
(CEF) ground state. The residual resistivity of these crystals features a
smeared step centered around 9\,T, the predicted crossing field for the
lowest CEF levels. The CEF contribution to the magnetoresistance has a
weaker-than-calculated dependence on the field direction, suggesting that
interactions omitted from the CEF model lead to avoided crossing in the
effective levels of the Pr$^{3+}$ ion. The dome-shaped magnetoresistance
observed for $x = 0$ and 0.05 cannot be reproduced by the CEF model, and
likely results from fluctuations in the field-induced antiferroquadrupolar
phase.
\end{abstract}

\pacs{74.25.Ha, 74.70.Tx}

\maketitle

\section{Introduction}

PrOs$_4$Sb$_{12}$, the first discovered Pr-based heavy fermion and
superconductor,\cite{Bauer} remains a focus of extensive theoretical and
experimental investigation. Its significance lies in the fact that the origin
of the heavy-fermion behavior is associated with non-Kramers $f$-electron ions,
for which the conventional Kondo effect seems unlikely. Our previous
specific-heat results in magnetic fields\cite{Rotundu} established that the
crystalline-electric-field (CEF) ground state is a nonmagnetic $\Gamma_1$
singlet. The field dependence of the CEF Schottky anomaly for fields greater
than 14\,T is clearly inconsistent with the alternative scenario of a
nonmagnetic $\Gamma_3$ doublet ground state. This conclusion was independent
of whether the exact $T_h$ point-group symmetry or the higher (approximate)
$O_h$ symmetry was assumed for the Pr sites.\cite{Takegahara}
The singlet nature of the CEF ground state was subsequently confirmed by
inelastic neutron scattering measurements and their analysis within the
$T_h$ symmetry scheme.\cite{Goremychkin}

Despite the overwhelming evidence in favor of a singlet CEF ground state,
there are experimental results for PrOs$_4$Sb$_{12}$ that seem to be
better understood in terms of a doublet ground state.
For example, the magnetoresistance\cite{Bauer,Frederick,Lacerda,Sugawara} at
1.4\,K exhibits a dome-like shape that is consistent with model calculations
of the CEF resistivity for a $\Gamma_3$ ground state and inconsistent with
similar calculations for a $\Gamma_1$ ground state.\cite{Frederick}
However, the CEF resistivity is a single-ion property that might be strongly
affected in PrOs$_4$Sb$_{12}$ by lattice coherence and by strong quadrupolar
and exchange interactions.
To probe this possibility, we have performed magnetoresistance measurements
on single-crystal Pr$_{1-x}$La$_x$Os$_4$Sb$_{12}$, in which lattice
translational symmetry is broken and intersite effects should be weaker than
in the pure compound.
Based on previously published magnetic susceptibility and specific heat
results,\cite{Rotundu2} we do not expect significant changes in CEF energies
(and eigenstates) of Pr upon doping with La. In addition, we have extended
magnetoresistance measurements of the undoped material down to 20\,mK.

We find that the magnetoresistance of pure PrOs$_4$Sb$_{12}$ at 20\,mK is
inconsistent with model calculations for either the $\Gamma_3$ or the
$\Gamma_1$ CEF ground state, and conclude that the dome feature most probably
results from fluctuations in the field-induced antiferroquadrupolar (AFQ)
phase. On dilution of the Pr lattice with La, the dome in the magnetoresistance
is replaced by a smeared step that is consistent with the picture of a
$\Gamma_1$ singlet CEF ground state but not with a $\Gamma_3$ doublet.
The dependence of the $f$-electron contribution to the magnetoresistance
on the direction of the magnetic field is smaller than is predicted
theoretically based on a CEF model. This discrepancy suggests that
interactions omitted from the CEF model lead to avoided crossing in the
effective levels of the Pr$^{3+}$ ion.

\section{Methods}

Results are presented below for Pr$_{1-x}$La$_x$Os$_4$Sb$_{12}$ with
four different La concentrations: $x=0$, 0.05, 0.3, and 0.67.
For $x=0$, 0.3, and 0.67, we grew large single crystals (cubes with masses
as large as 50\,mg) on which accurate magnetic susceptibility measurements
were performed up to 300\,K in order to extract the room-temperature
paramagnetic effective moment.
In each case, this moment was within 10\% of that expected for Pr$^{3+}$.
(For the undoped compound, this finding contradicts a wide range
of values reported in literature.)
The superconducting transition temperatures $T_c$ of the large single crystals
and of smaller resistivity bars, also obtained in the same growths, were
checked via ac susceptibility measurements. A good agreement between $T_c$
values of large and small crystals confirmed the stoichiometry assigned
to samples used in this study. The residual resistivity ratio (the ratio of
the resistance at room temperature to that extrapolated to $T=0$) was
RRR $= 100$, 50, 180, and 170 for $x=0$, 0.05, 0.3, and 0.7, respectively.
The value RRR $= 100$ exceeds those reported
previously\cite{Bauer,Sugawara,Sugawara2} for pure PrOs$_4$Sb$_{12}$,
indicative of the high quality of our samples.
The $x=0.05$ crystal was from the batch for which results were
reported in Ref.~\onlinecite{Rotundu2}.

The resistivity was measured by a conventional four-probe technique. The
estimated uncertainty in the determination of the absolute value of the
resistivity was 30\% due to the unfavorable geometry of the crystals.
Within this uncertainty, the resistivity at room temperature was the same
in all cases. In the plots below, we have scaled the resistivity
of each sample to a zero-field room-temperature value of 300\,$\mu\Omega$cm,
in the range reported previously. It is important to emphasize that this
scaling procedure is in no way essential for the conclusions of the paper,
which are based on the temperature and field dependence of the resistivity
of a given sample.

We calculated the CEF contribution \rhoCEF\ to the electrical
resistivity via the method applied by Fisk and Johnston\cite{Fisk} to the
resistivity of PrB$_6$ and by Frederick and Maple\cite{Frederick} to the
magnetoresistance of PrOs$_4$Sb$_{12}$.
This method focuses on a single Pr ion, neglects intersite effects, and takes
no account of the direction of the current relative to the crystal axes or to
the magnetic field.
Our calculations started with one or other of two forms for $\hat{H}_0$,
the CEF Hamiltonian for Pr$^{3+}$ in zero magnetic field:
that (corresponding in the $O_h$-symmetry notation of Ref.\ \onlinecite{Lea}
to $W=-2.97$\,K and $x=-0.7225$) deduced\cite{Frederick} by fitting the
temperature dependence of the zero-field resistivity of PrOs$_4$Sb$_{12}$;
or the Hamiltonian (described in the $T_h$-symmetry notation of
Ref.\ \onlinecite{Takegahara} by $W=3.0877$\,K, $x=0.45991$, and $y=0.10503$)
determined from elastic\cite{Kohgi} and inelastic\cite{Goremychkin} neutron
scattering.
Henceforth, we refer to these cases as the ``doublet'' and ``singlet'' CEF
scheme, respectively, according to the ground-state degeneracy of $\hat{H}_0$.

The CEF states in a magnetic field $\mathbf{H}$ were obtained by diagonalizing
the Hamiltonian $\hat{H}_0 + g \mu_B \mathbf{H}\cdot\hat{\mathbf{J}}$, where
$g = 4/5$ is the Land\'{e} $g$ factor for $Pr^{3+}$ and $\hat{\mathbf{J}}$ is
the $f$-electron angular momentum operator. The CEF resistivity is completely
determined by these CEF states, the temperature $T$, and two constants
$\rho_{\mathrm{ex}}$ and $\rho_A$, which parametrize, respectively, the overall
strengths of magnetic exchange and the aspherical Coulomb interaction between
the $4f$ and conduction electrons. Following Ref. \onlinecite{Frederick}, we
took $\rho_{\mathrm{ex}}=\rho_A = 0.25\,\mu\Omega$cm. However, our conclusions
are insensitive to the particular choice of constants.

\section{Results and Discussion}

\begin{figure}
\includegraphics[width=3.0in]{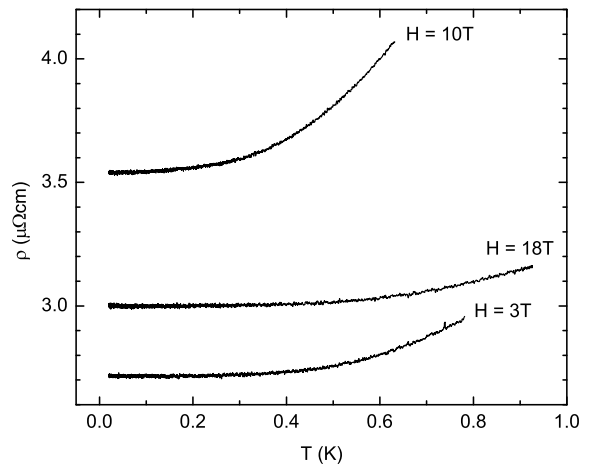}
\caption{\label{rho:low-T}%
Resistivity vs temperature for PrOs$_4$Sb$_{12}$ in three different magnetic
fields, with both current and field along the (001) direction.}
\end{figure}

Figure \ref{rho:low-T} shows the resistivity of undoped PrOs$_4$Sb$_{12}$ for
three representative fields, with both current and magnetic field oriented
along the (001) direction.
The results are similar to those reported by other groups.\cite{Bauer,Maple}
Below 200--300\,mK the resistivity saturates but has a strong field variation.
The residual resistivity $\rho_0$ can be obtained using the previously
noted\cite{Bauer,Maple} temperature variation at fixed field:
$\rho(T)=\rho_0+ B T^n$ with $n>2$.
Within the precision of our measurement, there is no difference between
$\rho_0$ obtained in this manner and $\rho(T=\mathrm{20\,mK})$.

\begin{figure}
\includegraphics[width=3.0in]{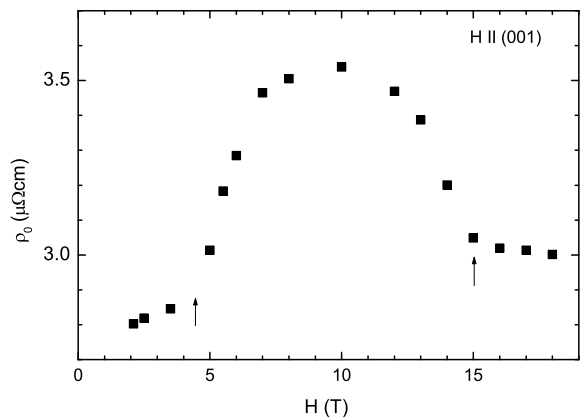}
\caption{\label{rho_0:pure}%
Residual resistivity vs magnetic field for PrOs$_4$Sb$_{12}$,
with both current and field along the (001) direction.
Arrows indicate boundaries between paramagnetic and
field-induced ordered phases.}
\end{figure}

\begin{figure}
\includegraphics[width=3.3in]{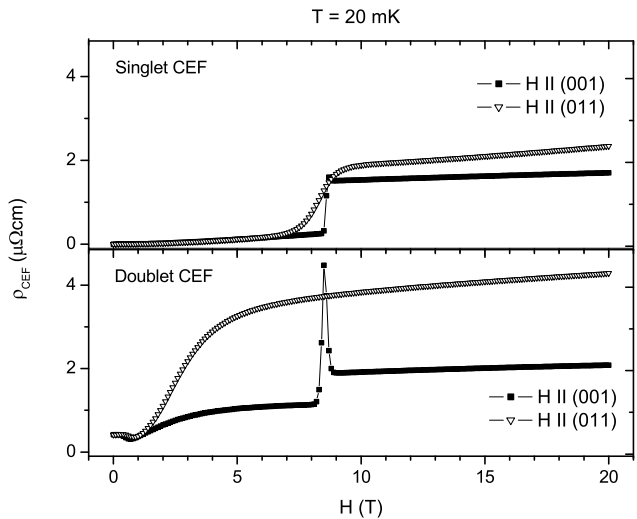}
\caption{\label{rho:2_schemes}%
Theoretical CEF resistivity at 20\,mK vs magnetic field calculated within the
singlet (upper panel) and doublet (lower panel) CEF schemes, for fields
along (001) ($\scriptstyle\blacksquare$) and (011) ($\triangledown$).}
\end{figure}

This residual resistivity (or resistivity at 20\,mK), when plotted against
magnetic field (Fig.\ \ref{rho_0:pure}), has a dome shape centered around
9--10\,T.
Such a dome-shaped magnetoresistance has been reported previously at the
somewhat higher temperatures of 1.4\,K (Ref.~\onlinecite{Frederick}) and
0.36\,K (Ref.~\onlinecite{Sugawara}). Two explanations for this dome have
been considered: field-induced long-range antiferroquadrupolar (AFQ) order,
and crossing of the lowest CEF levels. It is striking that $\rho_0(H)$ rises
sharply at the AFQ boundaries, indicated by arrows in Fig.\ \ref{rho_0:pure},
and peaks around 10\,T, where the AFQ transition temperature is highest.
However, Frederick and Maple have shown (see Fig.\ 2 of
Ref.\ \onlinecite{Frederick}) that the width, peak position, and height of
the dome in the magnetoresistance of PrOs$_4$Sb$_{12}$ at 1.4\,K are
reproduced quite well by the single-ion CEF resistivity computed within the
doublet scheme. (By contrast, \rhoCEF\ for the singlet scheme shows a step-like
increase in the vicinity of the crossing field at which the lowest $T_h$
$\Gamma_4^{(2)}$ level falls in energy below the $\Gamma_1$ singlet.)

We find that neither CEF scheme accounts satisfactorily for the
magnetoresistance measured at 20\,mK (Fig.~\ref{rho_0:pure}), which shows a
dome of similar width to that at 1.4\,K. Irrespective of the CEF scheme,
\rhoCEF\ for fields oriented along the (001) direction (square symbols in
Fig.~\ref{rho:2_schemes}) is discontinuous at the crossing field and
essentially flat at higher fields. It therefore seems that the low-temperature
magnetoresistance of PrOs$_4$Sb$_{12}$ is dominated by effects beyond those
considered in the single-ion CEF model.

\begin{figure}
\includegraphics[width=3.0in]{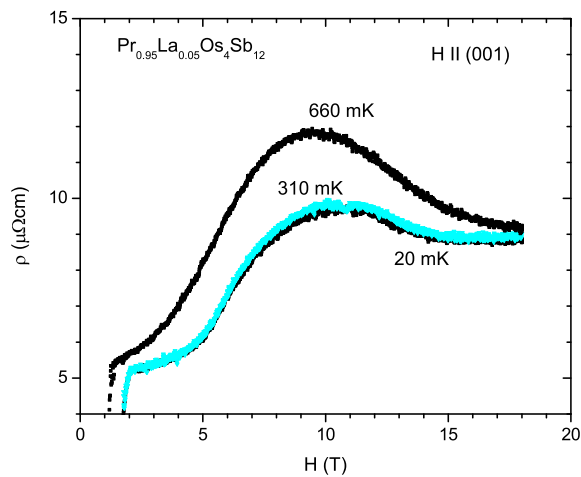}
\caption{\label{rho:x=0.05}%
(color online)
Longitudinal magnetoresistance of Pr$_{0.95}$La$_{0.05}$Os$_4$Sb$_{12}$ at
20, 310, and 660\,mK, for current and field along the (001) direction.}
\end{figure}

We now turn to the effects of La doping.
Figure \ref{rho:x=0.05} shows the magnetoresistance of
Pr$_{0.95}$La$_{0.05}$Os$_4$Sb$_{12}$ at 20, 310, and 660\,mK. Similarly to
PrOs$_4$Sb$_{12}$, there is a negligible temperature variation of the
resistivity below 300\,mK in fields above-critical for superconductivity
(as evidenced by the overlap of the 20-mK and 310-mK isotherms). However,
the shape of the dome for $x=0.05$ is much less symmetric about the peak field
than its $x=0$ counterpart.
Between 2\,T and 10\,T, $\rho$(T=20 mK) for the doped sample increases by
over 80\%, compared to a 25\% increase for the undoped material, whereas
the resistivity drop above 10\,T is greater in percentage terms for $x=0$.

\begin{figure}
\includegraphics[width=3.3in]{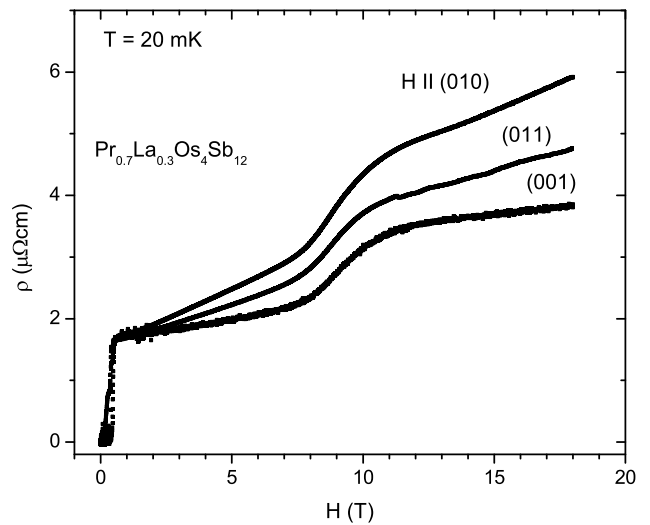}
\caption{\label{rho:x=0.3@20mK}%
Magnetoresistance of Pr$_{0.7}$La$_{0.3}$Os$_4$Sb$_{12}$ at 20mK for three
different orientations of the magnetic field. The current direction was (001).}
\end{figure}

The magnetoresistance becomes qualitatively different at higher La doping.
Figure \ref{rho:x=0.3@20mK} shows the 20-mK magnetoresistance of
Pr$_{0.7}$La$_{0.3}$Os$_4$Sb$_{12}$ for fields along (001), (011), and (010);
in each case, the current passed along the (001) direction.
All three isotherms exhibit a pronounced but wide step, centered near 9--10\,T,
superimposed on a linear background. In the investigated field range this
$x=0.3$ material does not exhibit the dome structure characteristic of $x=0$
and 0.05. For each curve, $\rho$ versus $H$ is approximately linear above
13\,T. The resistivity of the non-$f$-electron analog LaOs$_4$Sb$_{12}$,
measured at 0.36\,K, has a quite large and approximately linear field
dependence.\cite{Sugawara} Furthermore, the directional dependence of the
magnetoresistance of LaOs$_4$Sb$_{12}$---$d\rho/dH$ being larger along
(011) than along (001)---is in agreement with the trend of the linear background
in Fig. \ref{rho:x=0.3@20mK}. It thus seems that the differences between the
high-field slopes of $\rho(H)$ in Fig. \ref{rho:x=0.3@20mK} can be attributed
primarily to non-$f$-electron contributions to the magnetoresistance.
Subtracting such linear parts results in very similar curves (not shown) for
all three field directions. We conclude that the $f$-electron magnetoresistance
in this moderately doped material is nearly isotropic.

Compared to the cases $x=0$ and $x=0.05$, the low-temperature magnetoresistance
of Pr$_{0.7}$La$_{0.3}$Os$_4$Sb$_{12}$ for fields along (001) is much closer to
that given by the CEF model. The 20-mK measurements (Fig.\ \ref{rho:x=0.3@20mK})
are more consistent with the singlet CEF scheme than with the doublet scheme,
in that the latter predicts a sharp peak that is absent in the data. A second
and stronger argument in favor of the singlet scheme is provided by the
near-isotropy of the $f$-electron magnetoresistance noted in the previous
paragraph. Figure \ref{rho:2_schemes} plots the CEF resistivity for fields along
(001) and (011). The doublet scheme (lower panel in Fig.\ \ref{rho:2_schemes})
predicts a highly anisotropic \rhoCEF\ stemming from the fact that the lowest
two CEF levels cross at 8.5\,T along (001), but instead diverge along (011).
In the singlet CEF scheme (upper panel in Fig.\ \ref{rho:2_schemes}), the
anisotropy is much smaller because the lowest CEF levels cross at 8.6\,T along
(001), while along the (011) direction they anticross at 8.3\,T with a minimum
gap of only 0.7\,K. Unlike the doublet CEF scheme, the singlet scheme does a
reasonable job of reproducing the measured (011) magnetoresistance.
However, it underestimates the width of the step for fields along (001), and
hence still overestimates the anisotropy in the $f$-electron magnetoresistance
extracted from Fig.\ \ref{rho:x=0.3@20mK}.

\begin{figure}
\includegraphics[width=3.3in]{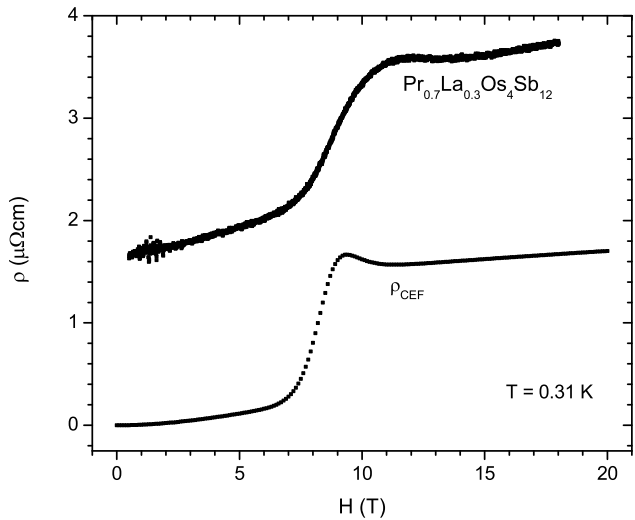}
\caption{\label{rho:x=0.3@310mK}%
Measured magnetoresistance of Pr$_{0.7}$La$_{0.3}$Os$_4$Sb$_{12}$ at
310\,mK for current and field along the (001) direction, and theoretical CEF
resistivity \rhoCEF\ for the same temperature and field direction.}
\end{figure}

\begin{figure}
\includegraphics[width=3.3in]{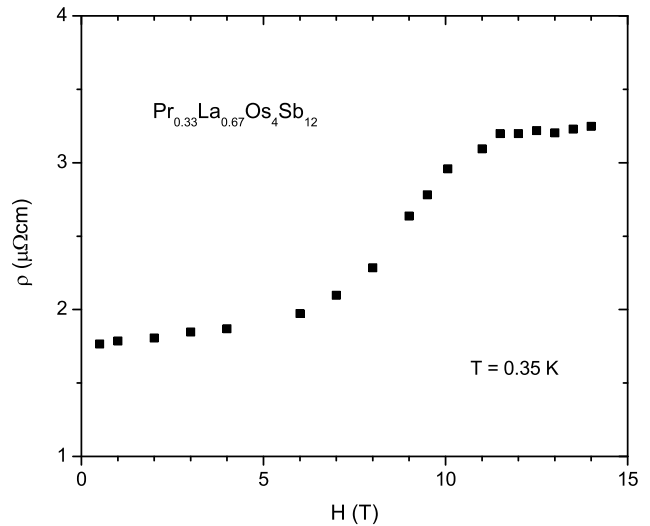}
\caption{\label{rho:x=0.67@350mK}%
Magnetoresistance of Pr$_{0.33}$La$_{0.67}$Os$_4$Sb$_{12}$ at
350\,mK for current and field along the (001) direction.}
\end{figure}

Figure \ref{rho:x=0.3@310mK} shows that at the higher temperature of 310\,mK,
there is much better agreement between the magnetoresistance of
Pr$_{0.7}$La$_{0.3}$Os$_4$Sb$_{12}$ along (001) and and \rhoCEF\ calculated
within the singlet CEF scheme. At this temperature, the thermal smearing
of the step in \rhoCEF\ matches quite well the width of the rise
in the measured magnetoresistance. The results of similar calculations for
the doublet scheme (not shown, but very similar to the 350-mK results in
Fig.\ 2 of Ref.\ \onlinecite{Frederick}) are in gross disagreement with the
measurement.

The character of the magnetoresistance seems to be little changed by further
La dilution. The longitudinal magnetoresistance for $x=0.67$ was investigated
down to 0.35\,K and in fields to 14\,T. The magnetoresistance at the lowest
temperature (Fig. \ref{rho:x=0.67@350mK}) exhibits essentially identical
magnetic field dependence to that for $x=0.3$.

The main difference between the $f$-electron magnetoresistance inferred for
dilute-Pr alloys and that calculated in the singlet CEF scheme relates to the
low-temperature width of the step along the (001) field direction. The CEF
model predicts an almost discontinuous jump of the magnetoresistance at 20\,mK
at the level-crossing field, while the rise in the measured magnetoresistance
of Pr$_{0.7}$La$_{0.3}$Os$_4$Sb$_{12}$ takes place over 3--4\,T.
Since there is better agreement between the measured and theoretical
magnetoresistance along the (011) direction, where level anticrossing is
expected, we speculate that interactions omitted from the CEF model mix the
lowest levels, preventing any crossing even along high-symmetry field
directions and thereby producing isotropic magnetoresistance.
These interactions are most likely nonlocal. We note, however, that a
mean-field treatment of intersite magnetic and quadrupolar interactions
between Pr ions did not find avoided crossing.\cite{Kohgi,Rotundu}

Figure \ref{rho:x=0.3@310mK} shows that the measured midpoint field for the
magnetoresistance rise in Pr$_{0.7}$La$_{0.3}$Os$_4$Sb$_{12}$ is about 1\,T
higher than is predicted based on the CEF level scheme determined
for pure PrOs$_4$Sb$_{12}$. This perhaps points to a small shift in the CEF
levels upon doping to 30\% La. The similarity between the magnetoresistance
steps observed for $x=0.3$ and 0.67 suggests that there is little further
evolution of the CEF energies over this doping range. A weak dependence of
CEF levels on La doping is in agreement with our specific-heat measurements
of the Schottky anomaly in lightly doped alloys,\cite{CEF} and is also
supported by the nearly invariant temperature of the maximum in the magnetic
susceptibility.\cite{Rotundu2}

The contrast between the resistivity versus field curves for the $x=0$ and
$x=0.3$ samples is striking. The obvious differences between these two
compositions are the presence of a field-induced ordered phase for the pure
material and the absence of translational symmetry in the diluted case.
Our previous specific heat measurements\cite{CEF} indicate that the
field-induced ordered (AFQ) phase disappears somewhere near $x=0.2$.
Since the model for the CEF resistivity are single-site in character,
it seems likely that the dome-shaped magnetoresistance observed for
$x=0$ and 0.05 is associated with the presence of long-range order
in these materials, perhaps through enhanced scattering of conduction
electrons caused by fluctuations in the AFQ order parameter.

In summary, we have shown that a simple CEF model accounts quite well
for the $f$-electron contribution to the magnetoresistance of
Pr$_{1-x}$La$_x$Os$_4$Sb$_{12}$ with $x = 0.3$ and 0.67. The weak dependence
of this contribution on field direction is consistent with the existence of a
singlet CEF in zero magnetic field, with avoided level crossing in applied
fields. At the lowest temperatures, the magnetoresistance for $x=0$ and
$x=0.05$ is inconsistent with the results of CEF model calculations.
This discrepancy is attributed to the long-range order present in the
pure and lightly doped materials.

\begin{acknowledgments}

Stimulating discussions with P.\ Kumar are acknowledged.
This work has been supported by U.S. Department of Energy (DOE) Grant
No.\ DE-FG02-99ER45748, by National Science Foundation (NSF) Grant No.\
DMR-0312939, and by the National High Magnetic Field Laboratory.
\end{acknowledgments}

\end{document}